# Micromechanical study of the dilatational response of porous solids with pressure-insensitive matrix displaying tension-compression asymmetry

J. L. Alves[1] and Oana Cazacu[2*]

[1]CT2M -Department of Mechanical Engineering, University of Minho, Portugal.

[2]Department of Mechanical and Aerospace Engineering, University of Florida, REEF, 1350 N. Poquito Rd., Shalimar, FL 32579, USA.

**Abstract**

In this paper, the dilatational response of porous solids with pressure-insensitive matrix displaying strength differential (SD) effects is investigated. To this end, micromechanical finite-element analyses of three-dimensional unit cells are carried out. The matrix behavior is governed by the isotropic form of Cazacu et al. (2006) criterion that accounts for SD effects through a parameter k. Simulation results are presented for axisymmetric tensile loadings corresponding to fixed values of the stress triaxiality for the two possible values of the Lode parameter, $\mu_\Sigma$. For moderate and high stress triaxialities, it is shown that for materials for which the matrix tensile strength is larger than its compressive strength (k > 0), under tensile loadings corresponding at $\mu_\Sigma =1$ the void growth rate is much faster than in the case of tensile loadings at $\mu_\Sigma = -1$. The opposite holds true for materials with matrix tensile strength lower than its compressive strength (k< 0). This drastic difference in porosity evolution is explained by the distribution of the local plastic strain and stresses, which are markedly different than in a von Mises material (i.e. no SD effects of the matrix).

[*] Corresponding author: Tel: +1 850 833 9350; Fax: +1 850 833 9366.
E-mail address: cazacu@reef.ufl.edu

# 1. Introduction

Ductile failure in metals occurs due to the presence of defects such as voids and microcracks (McClintock, 1968). The ability to accurately describe the evolution of voids in a ductile metal is essential for predicting its failure. Beginning with the pioneering studies of Needleman (1972), Tvergaard (1981), Koplik and Needleman (1988), micromechanical finite-element (FE) analyses of unit cells have been used to provide fundamental understanding of the mechanical response of porous solids (e.g. Richelsen and Tvergaard, 1994, Zhang et al., 2001; Srivastava and Needleman, 2012, Alves et al., 2014, etc.). In all these micromechanical studies, it was assumed that the plastic flow of the matrix (void-free material) is governed by the von Mises criterion.

Finite-element cell model calculations as well as analytical criteria for porous solids with matrix having a different response in tension and compression i.e. displaying strength differential (SD) effects have also been proposed. For example, for the case when the matrix obeys the Drucker-Prager pressure-sensitive criterion and associated flow rule, FE calculations of the yield stresses of the porous solid were reported by Trillat et al. (2006) while analytical criteria were developed by Barthelemy and Dormieux (2004), Guo et al. (2008), etc. It is to be noted that in all these models, the matrix plastic behavior is governed by a criterion that depends on the mean stress. Therefore, plastic behavior of the matrix is accompanied by volume changes.

However, fully-dense hexagonal close packed (HCP) metallic materials display SD effects, although their plastic response is independent of mean stress (pressure-insensitive). In these materials, SD effects in plastic flow are related to the polarity of the plastic deformation mechanisms (crystallographic twinning) being operational at single crystal level (see Hosford and Allen (1973), Hosford, 1993; for experimental evidence of the correlation between deformation twinning and tension-compression asymmetry, see Khan et al. 2011 for Mg alloy AZ31B; for Ti materials, see Nixon et al., 2010; Gilles et al., 2011; Knezevic et al., 2013). Concerning modeling of deformation twinning and its effects on the texture evolution of HCP polycrystals within the framework of crystal plasticity, the reader is referred to the seminal work of Van Houtte and collaborators, e.g.

Van Houtte et al., 1978; Leffers and Van Houtte, 1989; Philipe et al., 1995; Coghe et al., 2012).

The fact that fully-dense metals with cubic crystal structure display SD effects although the plastic flow is pressure-insensitive was explained by Asaro and Rice (1977) as being due to small deviations from the classical Schmid law.

Few studies have been devoted to the study of porous solids with incompressible matrix displaying SD effects. Using a kinematic limit-analysis approach, Cazacu and Stewart (2009) developed a plastic potential for porous solids with matrix governed by Cazacu et al. (2006) criterion that is pressure-insensitive, yet it accounts for SD effects.

According to Cazacu et al. (2009) model for porous solids, even very small tension-compression asymmetry of the matrix has a strong influence on yielding and porosity evolution. In contrast to the case when the matrix is described by the von Mises yield criterion, irrespective of loading yielding of the porous solid is strongly influenced by the third-invariant of the stress deviator, $J_3$.

Using Cazacu and Stewart (2009) plastic potential, Revil-Baudard and Cazacu (2013) analyzed the porosity evolution and the location of the zone corresponding to maximum porosity in notched specimens loaded in uniaxial tension. The FE simulations show that for materials for which the matrix uniaxial tensile strength is larger than its uniaxial compressive strength, void growth and porosity distribution are similar to that in porous materials with von Mises matrix. On the other hand, for porous materials for which the matrix tensile strength is lower than its compressive strength, the void growth rate is much slower, and if the strength differential is pronounced the location of the maximum damage zone shifts from the center to the surface of the specimen.

In this paper, we investigate the influence of the particularities of the plastic flow of the incompressible matrix, namely its tension-compression asymmetry, on the evolution of porosity, and how it affects the ductility of the porous material by conducting a micromechanical FE study. The porous medium is represented as a three-dimensional (3-D) regular spatial array of initially spherical voids packed in a fully dense matrix. The matrix is considered to be elastic-plastic, the plastic response being described by the isotropic form of Cazacu et al. (2006)'s yield criterion. This yield criterion is pressure-insensitive. It involves all principal values of the stress deviator (or equivalently depends on both invariants of the stress deviator), and a scalar material parameter, k, which is related to SD effects in plastic flow.

The outline of the paper is as follows. We begin with a brief presentation of the isotropic form of Cazacu et al. (2006) yield criterion. Next, the micromechanical unit-cell model and the method of analysis are discussed. In Section 3-4 are presented simulation results for porous materials with incompressible matrix characterized by different SD ratios. For each porous material, the macroscopic tensile loadings imposed are such that the principal values of the applied stress, $\Sigma_1$, $\Sigma_2$, $\Sigma_3$ follow a prescribed proportional loading history corresponding to a constant stress triaxiality $T_\Sigma$. Specifically, the response is investigated under axisymmetric loadings ($\Sigma_1=\Sigma_2$) where the axial overall stress $\Sigma_3$ is adjusted so that a fixed value of the stress triaxiality is maintained. To investigate the effects of the third-invariant $J_3$, for any given specified value of the stress triaxiality, void growth is analyzed for loadings corresponding to either (a) $\Sigma_3>\Sigma_1=\Sigma_2$ ($J_3>0$ for the entire loading history) or (b) $\Sigma_3<\Sigma_1=\Sigma_2$ ($J_3<0$). Detailed analyses of the distribution of local stresses and local plastic strains are presented for a moderate value of the triaxiality $T_\Sigma=1$ (Sections 3-4). It is shown that for the same imposed loading, the distribution of the local plastic strain and local mean stress in the porous material is strongly influenced by the value of the material parameter k, which accounts for SD effects. This in turn affects any aspect of porosity evolution. The main findings of this study and concluding remarks are given in Section 5.

## 2. Problem Formulation and Method of Analysis

### 2.1. Isotropic form of Cazacu et al., 2006 yield criterion

The isotropic form of Cazacu et al. (2006) yield criterion will be used to model the plastic behavior of the matrix. This criterion is only briefly presented in what follows, while a detailed description can be found in Cazacu et al. (2006). Application /verification/validation of this model for a variety of loading conditions such as uniaxial tension, uniaxial compression, bending, torsion can be found for example in Plunkett et al. (2006), Cazacu et al. (2012; 2014), etc. The isotropic form of Cazacu et al. (2006) criterion is:

$$\sqrt{\frac{9}{2(3k^2-2k+3)}\sum_{i=1}^{3}\left(\left|\sigma'_i\right|-k\sigma'_i\right)^2} = \sigma_T, \qquad (1)$$

where $\sigma'_1, \sigma'_2, \sigma'_3$ are the principal values of the deviator of the Cauchy stress tensor, $\boldsymbol{\sigma}' = \boldsymbol{\sigma} - \sigma_m \mathbf{I}$; with $\sigma_m = \text{tr}(\boldsymbol{\sigma})/3$ denoting the mean stress and $\mathbf{I}$ the second-order identity tensor, while $\sigma_T$ is the yield stress in uniaxial tension. The only material parameter involved in the criterion is the parameter k, which is intimately linked with specific single-crystal plastic deformation mechanisms (e.g. see Lebensohn and Cazacu, 2012).

Using Eq. (1), it can be easily shown that this parameter depends solely on the ratio $\sigma_T/\sigma_C$ between the uniaxial flow stresses in tension and compression, respectively (for more details, see Cazacu et al. (2006)), i.e.

$$k = \frac{1-h}{1+h}, \text{ where } h = \sqrt{\frac{2-(\sigma_T/\sigma_C)^2}{2(\sigma_T/\sigma_C)^2-1}} \qquad . \qquad (2)$$

If a material does not display tension-compression asymmetry, i.e. the yield stress in uniaxial tension, $\sigma_T$, is the same as the yield stress in uniaxial compression, $\sigma_C$, the parameter k = 0, and the isotropic form of Cazacu et al. (2006) criterion reduces to the von Mises criterion (see Eq. (1)). If a material displays SD effects then k ≠ 0, which means that the plastic flow depends on the sign and ordering of all principal values of $\boldsymbol{\sigma}'$ or alternatively on both invariants of the stress deviator (see Eq. (1)). Specifically, for a material with $\sigma_T/\sigma_C < 1$, k < 0 while for a material with $\sigma_T/\sigma_C > 1$, k > 0. As an example, in Fig.1 (a) are shown the projections in the octahedral plane (plane normal to the hydrostatic axis, $\sigma_1 = \sigma_2 = \sigma_3$) of the yield surface given by Eq. (1) for k = - 0.3 ($\sigma_T/\sigma_C =$ 0.83) in comparison with the von Mises yield surface k = 0 ( von Mises: $\sigma_T/\sigma_C = 1$), while in Fig 1(b) is shown a comparison between the yield surface corresponding to k = 0.3 ($\sigma_T/\sigma_C$ = 1.21), and von Mises respectively. Note that for k different from zero there is strong dependence of the yield loci on $J_3$ as evidenced by their shape (i.e. triangles with rounded corners while the projection of the von Mises yield criterion is a circle).

**2.2. Unit-cell model**

It is assumed that the porous polycrystal contains a regular and periodic 3-D array of initially spherical voids. The inter-void spacing is considered to be the same in any direction. The unit cell, which takes into account the periodicity of the porous medium, is initially cubic with side lengths 2 $C_0$ and contains a single spherical void of radius $r_0$ at its center. Thus, the initial porosity is:

$$f_0 = \frac{\pi}{6}\left(\frac{r_0}{C_0}\right)^3. \tag{3}$$

Cartesian tensor notations are used and the origin of the coordinate system is taken at the center of the void (see Fig. 2 (a)). Let **u** denote the incremental displacement between the current and reference configuration, and **t** the prescribed Cauchy stress vector, defined on the current configuration. Symmetry conditions are imposed on the planes x = 0, y = 0, and z = 0, respectively:

$$\begin{aligned}
u_1(0, y, z) &= 0, & t_2(0, y, z) &= 0, \ t_3(0, y, z) = 0, \\
u_2(x, 0, z) &= 0, & t_1(x, 0, z) &= 0, \ t_3(x, 0, z) = 0, \\
u_3(x, y, 0) &= 0, & t_1(x, y, 0) &= 0, \ t_2(x, y, 0) = 0.
\end{aligned} \tag{4}$$

Therefore, only one eight of the unit cell needs to be analyzed numerically (see Fig. 2 (b)). To simulate the constraints of the surrounding material, we enforce that the faces of the unit cell, which are initially planes parallel to the coordinate planes, remain planes and are shear free. The boundary conditions imposed on the faces of the unit cell are:

$$u_1(C_0, y, z) = U_1^*(t), \quad t_2(C_0, y, z) = t_3(C_0, y, z) = 0,$$

$$u_2(x, C_0, z) = U_2^*(t), \quad t_1(x, C_0, z) = t_3(x, C_0, z) = 0, \quad (5)$$

$$u_3(x, y, C_0) = U_3^*(t), \quad t_1(x, y, C_0) = t_2(x, y, C_0) = 0.$$

The void is considered to be traction-free. The macroscopic true stresses $\Sigma_1$, $\Sigma_2$, $\Sigma_3$ are defined as:

$$\Sigma_1 = \frac{1}{C_2 C_3} \int_0^{C_2} \int_0^{C_3} t_1 \, dzdy, \quad \Sigma_2 = \frac{1}{C_1 C_3} \int_0^{C_3} \int_0^{C_1} t_2 \, dzdx,$$

$$\Sigma_3 = \frac{1}{C_1 C_2} \int_0^{C_1} \int_0^{C_2} t_3 \, dxdy, \quad (6)$$

where $C_i = C_0 + U_i^*$ are the current cell dimensions. The porous material being isotropic, its mechanical response is fully characterized by the isotropic invariants of the overall stress, i.e.:

$$\Sigma_m = \frac{1}{3}(\Sigma_1 + \Sigma_2 + \Sigma_3); \quad \Sigma_e = \sqrt{3J_2^{\Sigma'}} = \sqrt{\frac{3}{2}(\Sigma_1'^2 + \Sigma_2'^2 + \Sigma_3'^2)}; \quad J_3 = \Sigma_1' \Sigma_2' \Sigma_3', \quad (7)$$

where $\Sigma_i' = \Sigma_i - \Sigma_m$, $i = 1...3$. In the analysis, we will refer to the following combinations of the stress invariants: the von Mises stress triaxiality ratio, $T_\Sigma$, and the Lode parameter $\mu_\Sigma$ (defined in terms of stress invariants by Drucker, 1949):

$$T_\Sigma = \Sigma_m / \sqrt{3J_2^\Sigma}$$

$$\mu_\Sigma = \frac{3\sqrt{3}}{2} \cdot \frac{J_3^\Sigma}{\left(J_2^\Sigma\right)^{3/2}}. \quad (8)$$

The time histories of the displacements, $U_1^*(t)$, $U_2^*(t)$, and $U_3^*(t)$ in Eq. (5) are determined by the analysis in such a way that the macroscopic Cauchy stresses $\Sigma_1$, $\Sigma_2$, $\Sigma_3$ follow a prescribed proportional loading given by:

$$\frac{\Sigma_1}{\Sigma_2} = 1 \text{ (axisymmetric loadings)} \quad \text{and} \quad \frac{\Sigma_1}{\Sigma_3} = \rho, \tag{9}$$

where $\rho$ is a prescribed constant such as to ensure that the stress triaxiality ratio, $T_\Sigma$, has a fixed specified value for the entire deformation process. The overall (macroscopic) principal strains and the macroscopic von Mises equivalent strain $E_e$ are calculated as follows:

$$E_1 = \ln\left(\frac{C_1}{C_0}\right), \quad E_2 = \ln\left(\frac{C_2}{C_0}\right); \quad E_3 = \ln\left(\frac{C_3}{C_0}\right)$$

$$E_e^2 = \frac{2}{3}\left(E_1^2 + E_2^2 + E_3^2\right), \tag{10}$$

where $C_0$ and $C_i$, $i = 1...3$ are the initial and current cell dimensions.

We conclude the presentation of the FE unit cell model, with a comment concerning the FE implementation of the above boundary value problem. In the FE implementation, the degrees of freedom of all FE nodes belonging to the same planar bounding surface of the cubic surface are associated in the global stiffness matrix, and the equations of all these degrees of freedom are replaced by only one unknown variable. In this way, it is ensured that all initially planar boundary surfaces remain strictly flat during the entire loading history. Additionally, for each time increment and for all equilibrium cycles, the three imposed macroscopic forces on each planar bounding surface of the cubic cell are continuously updated in order to ensure the specified macroscopic Cauchy stress ratios on the final equilibrated configuration; the macroscopic non-equilibrated forces are introduced in the fully-implicit Newton-Raphson algorithm in order to improve its convergence rate. Finally, a convergence criterion imposes that, for each planar surface, the ratio between the norm of the difference between the prescribed and effective macroscopic forces and the norm of the prescribed macroscopic force must be smaller than 0.001.

As mentioned, we focus on the study of void growth under axisymmetric tensile loadings, i.e. $\Sigma_i > 0$ (see Eq. (6)). At the end of each time increment, the condition of constant proportionality between the true stresses (Eq.(9)) is strictly verified, so it is ensured that the macroscopic stress triaxiality, $T_\Sigma$, remains constant throughout the given deformation history. We will analyze void growth for the following triaxialities values : $T_\Sigma = 2/3, 1, 2$. For each specified value of the stress triaxiality, calculations are conducted for overall axisymmetric states corresponding to the two possible values of the Lode parameter, i.e. $\mu_\Sigma = 1$ and $\mu_\Sigma = -1$. The Lode parameter value $\mu_\Sigma = 1$ corresponds to axisymmetric loadings such that $\Sigma_3 > \Sigma_1 = \Sigma_2$ ($J_3 > 0$) while $\mu_\Sigma = -1$ corresponds to $\Sigma_3 < \Sigma_1 = \Sigma_2$ ($J_3 < 0$). In one case, two principal values of the stress deviator $\Sigma'$ are *compressive (negative)*, but the maximum principal value is *tensile (positive)* while for the other case two principal values of $\Sigma'$ are *tensile* (positive), but *the minor principal value, which is compressive (negative),* has the largest absolute value. The void volume fraction, f , is evaluated at the end of each time increment as:

$$f = \frac{V_{cell} - V_{matrix}}{V_{cell}}.$$

In the above equation, $V_{cell} = C_1 C_2 C_3$, where $C_i$ denote the current dimensions of the cell, while the volume of the deformed matrix, $V_{matrix}$, is determined directly from the integration of the FE domain using the FE formulation ($V_{matrix} = \sum_{i=1}^{N_E} V_i$, where $V_i$ is the volume of the element $i$ and $N_E$ is the total number of finite elements in the mesh).

Using the FE cell model, the response of three porous materials with matrix obeying the isotropic form of Cazacu et al. (2006) criterion (Eq. (1)) and associated flow rule will be investigated. Specifically, three porous materials characterized by different tension-compression asymmetry ratios, $\sigma_T/\sigma_C$, but having the same initial void volume fraction, $f_0$ = 0.0104 (which corresponds to $r_0/C_0 = 0.271$) will be considered. The respective SD ratios are: $\sigma_T/\sigma_C = 0.83$; $\sigma_T/\sigma_C = 1$ (von Mises material); and $\sigma_T/\sigma_C = 1.21$. Thus, these materials are characterized by k = -0.3, k=0, and k= +0.3 (see Eq. (1)-(2)).

In all the computations, all the other input material parameters are kept the same, i.e. the elastic properties (E= 200 GPa, ν = 0.33, where E is the Young modulus and ν is the Poisson coefficient) and the material parameters involved in the isotropic hardening law describing the evolution of the matrix tensile yield strength with local equivalent plastic strain, $\bar{\varepsilon}^p$, i.e.

$$Y = A(\varepsilon_0 + \bar{\varepsilon}^p)^n, \tag{11}$$

where Y is the current matrix tensile flow stress, A , n and $\varepsilon_0$ are material parameters. The numerical values of these parameters are: $Y_0$= 400 MPa, A= 881.53 MPa, n = 0.1, $\varepsilon_0$ = 0.00037. It follows that all the differences in behavior between the porous materials are due solely to SD effects of the matrix, which are described by the parameter k. The calculations are continued until either one of the two conditions are met: (i) relative void volume fraction $f/f_0$ >9 or, (ii) a macroscopic effective strain Ee =1 was reached.

The FE analyses were performed with DD3IMP (Menezes and Teodosiu, 2000, Oliveira et al. 2008), an in-house quasi-static elastoplastic FE solver with a fully-implicit time integration scheme. Figure 2(b) shows the initial FE mesh of one-eighth of the unit cubic cell consisting of 12150 elements (8-node hexahedral finite elements; selective reduced integration technique, with 8 and 1 Gauss points for the deviatoric and volumetric parts of the velocity field gradient, respectively) and a total of 13699 nodes. A mesh refinement study was carried out to ensure that the results are mesh-independent.

## 3. Analysis of the porosity evolution and its effects on the ductility of porous materials for axisymmetric tensile loading at $T_\Sigma =1$

Using the FE unit-cell model, we first examine the porosity evolution and its effects on the mechanical response of the porous materials for axial tensile loading corresponding to $T_\Sigma =1$ and $\mu_\Sigma =1$ ($\rho = \Sigma_3/\Sigma_1 =2.5$ and $J_3 > 0$ during the entire loading history). This stress triaxiality corresponds to tensile loading of blunt notched specimens (e.g. see Tvergaard and Needleman, 1984). Fig. 3 shows a comparison between the macroscopic effective stress-macroscopic effective strain ($\Sigma_e$ vs. $E_e$) curves of the three porous materials; Fig. 4 shows the evolution of the void volume fraction, f, while Fig. 5 depicts the rate of void growth as a function of the macroscopic effective strain $E_e$. It is clearly seen that all aspects of the macroscopic response of the porous materials are influenced by the specificities of the plastic flow of the matrix described by the parameter k. Specifically, the maximum effective stress that is reached, and the maximum strain (i.e. the material's ductility) strongly depend on k (see Fig. 3). It is worth noting that the porous material characterized by k = -0.3 ($\sigma_T/\sigma_C = 0.83$) has the highest ductility, but the stress drop is also the most abrupt. This indicates that for this material failure is more catastrophic than in a porous von Mises material (k = 0) or in the porous material with matrix characterized by k = 0.3 ($\sigma_T/\sigma_C = 1.21$). The same conclusion can be drawn by examining the void volume fraction evolution (Fig. 4), and the rate of void growth (Fig. 5). Void growth is fastest in the material with k =+0.3 and slowest in the material with k =-0.3. It is worth noting that for the material with k = -0.3 the rate of void growth is almost constant for most of the deformation process. In contrast, for the porous von Mises material (k = 0) and for the porous material characterized by k = 0.3, damage accumulation is more gradual.

To detect the onset of void coalescence we use the procedure outlined in Srivastava and Needleman (2012), which is based on monitoring the evolution of the relative inter-void ligament size. Since for an isotropic material, the greatest reduction in ligament size occurs in the direction of minimum applied stress, the evolution of the relative inter-void ligament size in the x direction i.e. $l_1^r = l_1/l_1^0 = (C_1 - r_1)/(C_1^0 - r_0)$ with the macroscopic effective strain $E_e$ was examined (Fig. 6(a)). The evolution of the normalized cell dimension in the same direction is shown in Fig. 6(b). Note that for all materials, there is a decrease in $l_1^r$; and at a certain value of $E_e$ there is a change in slope. For a nearly rate-independent von Mises material (n =0.1) and same initial porosity, and loading history ($T_\Sigma =1$ and $\mu_\Sigma =1$), Koplik and Needleman (1988) reported FE results obtained using a cylindrical unit cell. It was shown that the onset of void coalescence is associated with strain localization in the ligament between adjacent voids leading to an overall uniaxial straining mode. The same was observed in the calculations here for the von Mises material (k =0), the transverse strain being almost constant ($dC_1/dE_e \approx 0$) beyond a critical value of $E_e$, marking coalescence (see Fig. 6(b)).

It is worth noting that the SD ratio of the matrix affects all aspects of damage evolution. For the porous material characterized by k = -0.3 ($\sigma_T/\sigma_C$ = 0.83), the rate of void growth is significantly lower (see Fig. 3-4) than in the other two materials, and the macroscopic strain $E_e$ at coalescence (strain localization) is more than three times higher than in the material with k = + 0.3 (see Fig. 6). Of the three materials, in the material with k = + 0.3, for which $\sigma_T > \sigma_C$, void growth is the fastest and consequently the decrease in relative ligament size is most rapid. This explains that for this loading, this material has much more reduced ductility than the other two materials (see also Fig. 3).

To better understand the reasons for the very strong difference in the rate of void growth between the three porous materials, we compare the local state fields corresponding to the same level of macroscopic true strain $E_e$ = 0.15. Note that this strain level corresponds to the early stages of the deformation process where macroscopically only a very slight difference between the stress-strain response of the three materials can be observed (see Fig. 3). Fig. 7 shows the contours of constant local equivalent plastic strain, $\bar{\varepsilon}^p$, corresponding to a macroscopic effective strain $E_e$ = 0.15. The local equivalent plastic strain $\bar{\varepsilon}^p$ is that associated with the effective stress associated with Cazacu et al. (2006) criterion given by Eq. (1), i.e. to $\tilde{\sigma}_e = \sqrt{\frac{9}{2(3k^2 - 2k + 3)} \sum_{i=1}^{3} (|\sigma'_i| - k\sigma'_i)^2}$ using the work-equivalence principle (Hill, 1987). The expression of $\dot{\bar{\varepsilon}}^p$ is:

$$\dot{\bar{\varepsilon}}^p = \begin{cases} \dfrac{1}{(1-k)} \cdot \sqrt{\dot{\varepsilon}_1^2 + \dot{\varepsilon}_2^2 + \left[\dfrac{3k^2 - 10k + 3}{3k^2 + 2k + 3}\right]\dot{\varepsilon}_3^2}, & \text{if } \dfrac{\dot{\varepsilon}_3}{\sqrt{\dot{\varepsilon}_1^2 + \dot{\varepsilon}_2^2 + \dot{\varepsilon}_3^2}} \leq \dfrac{-(3k^2 + 2k + 3)}{\sqrt{6(k^2 + 3)(3k^2 + 1)}} \\[2ex] \dfrac{1}{(1+k)} \sqrt{\dot{\varepsilon}_1^2 + \dot{\varepsilon}_2^2 + \left[\dfrac{3k^2 + 10k + 3}{3k^2 - 2k + 3}\right]\dot{\varepsilon}_3^2}, & \text{if } \dfrac{\dot{\varepsilon}_3}{\sqrt{\dot{\varepsilon}_1^2 + \dot{\varepsilon}_2^2 + \dot{\varepsilon}_3^2}} \geq \dfrac{3k^2 - 2k + 3}{\sqrt{6(k^2 + 3)(3k^2 + 1)}} \\[2ex] \dfrac{1}{(1-k)} \cdot \sqrt{\dot{\varepsilon}_2^2 + \dot{\varepsilon}_3^2 + \left[\dfrac{3k^2 - 10k + 3}{3k^2 + 2k + 3}\right]\dot{\varepsilon}_1^2}, & \text{if } \dfrac{\dot{\varepsilon}_1}{\sqrt{\dot{\varepsilon}_1^2 + \dot{\varepsilon}_2^2 + \dot{\varepsilon}_3^2}} \leq \dfrac{-(3k^2 + 2k + 3)}{\sqrt{6(k^2 + 3)(3k^2 + 1)}} \\[2ex] \dfrac{1}{(1+k)} \sqrt{\dot{\varepsilon}_2^2 + \dot{\varepsilon}_3^2 + \left[\dfrac{3k^2 + 10k + 3}{3k^2 - 2k + 3}\right]\dot{\varepsilon}_1^2}, & \text{if } \dfrac{\dot{\varepsilon}_1}{\sqrt{\dot{\varepsilon}_1^2 + \dot{\varepsilon}_2^2 + \dot{\varepsilon}_3^2}} \geq \dfrac{3k^2 - 2k + 3}{\sqrt{6(k^2 + 3)(3k^2 + 1)}} \\[2ex] \dfrac{1}{(1-k)} \cdot \sqrt{\dot{\varepsilon}_3^2 + \dot{\varepsilon}_1^2 + \left[\dfrac{3k^2 - 10k + 3}{3k^2 + 2k + 3}\right]\dot{\varepsilon}_2^2}, & \text{if } \dfrac{\dot{\varepsilon}_2}{\sqrt{\dot{\varepsilon}_1^2 + \dot{\varepsilon}_2^2 + \dot{\varepsilon}_3^2}} \leq \dfrac{-(3k^2 + 2k + 3)}{\sqrt{6(k^2 + 3)(3k^2 + 1)}} \\[2ex] \dfrac{1}{(1+k)} \sqrt{\dot{\varepsilon}_3^2 + \dot{\varepsilon}_1^2 + \left[\dfrac{3k^2 + 10k + 3}{3k^2 - 2k + 3}\right]\dot{\varepsilon}_2^2}, & \text{if } \dfrac{\dot{\varepsilon}_2}{\sqrt{\dot{\varepsilon}_1^2 + \dot{\varepsilon}_2^2 + \dot{\varepsilon}_3^2}} \geq \dfrac{3k^2 - 2k + 3}{\sqrt{6(k^2 + 3)(3k^2 + 1)}} \end{cases}$$

with $\dot{\varepsilon}_1$, $\dot{\varepsilon}_2$, $\dot{\varepsilon}_3$, being the principal values of the local plastic strain-rate tensor (for details of the derivation, see Cazacu et al. (2010)). Obviously, for k =0 (von Mises material), $\bar{\varepsilon}^p$ reduces to the classic von Mises equivalent strain.

The white regions in this Fig. 7 mark the elastic zones.

*It is worth noting that only in the material with k = -0.3, the entire domain (cell) is plastified.* However, for the porous von Mises material and the porous material characterized by k = + 0.3 ($\sigma_T/\sigma_C$ = 1.21), there exists a zone in the vicinity of the void along the vertical axis of the cross-section (Oz) where yielding did not occur. Specifically, for the porous von Mises material (k=0) the elastic zone is contiguous to the void while for the material characterized by k = 0.3, the elastic zone is slightly shifted upwards from the void.

Examination of the isocontours of the local equivalent plastic strain shows very marked differences in terms of the heterogeneity of plastic deformation and the distribution of the plastic zones within the domain. Note that for the porous material with matrix characterized by k = -0.3, the plastic deformation is more homogeneous than in the other materials. In contrast, at the same level of the macroscopic equivalent strain ($E_e$), in the porous von Mises material (k = 0) and in the porous material with matrix characterized by k = 0.3, the local plastic strain gradients are much stronger. The highest levels of local plastic deformation and most heterogeneity are found in the material characterized by k = + 0.3. A measure of the heterogeneity in plastic deformation within the domain is the ratio between the maximum local plastic strain in the entire domain, $\bar{\varepsilon}^p_{max}$, and the average of the local plastic strain, $\langle \bar{\varepsilon}^p \rangle$ defined as:

$$\langle \bar{\varepsilon}^p \rangle = \frac{1}{V} \int_V \bar{\varepsilon}^p dV \quad .$$

The highest is the ratio $\bar{\varepsilon}^p_{max} / \langle \bar{\varepsilon}^p \rangle$, the most heterogeneity there is. For the materials with k = +0.3, k=0, and k = -0.3, these ratios are 5.95, 5.4, and 4.16, respectively. Again the highest ratio (highest heterogeneity) is observed in the material with k = + 0.3 while the lowest is in the material with k = -0.3 ($\sigma_T/\sigma_C$ = 0.83). Thus, the local heterogeneity is always the lowest in the material with matrix characterized by k = -0.3.

Moreover, the distribution of the local stresses is markedly different depending on the SD ratio of the matrix, i.e. the value of k. Isocontours of the local normalized mean stress $\sigma_m/Y_0$ for each material are shown in Fig. 8. Note that in the porous material with matrix characterized by k = + 0.3 ($\sigma_T/\sigma_C$ = 1.21), the local mean stress, $\sigma_m/Y_0$, is positive in the entire domain while in the material with k = -0.3 ($\sigma_T/\sigma_C$ = 0.83) which is fully plastified, zones of negative (compressive) mean stress develop near the void. *As a consequence, for the latter material void growth is slowed down as compared to the porous material with matrix characterized by k = + 0.3. This correlates very well with the results presented in Fig. 3-5, in particular it explains the drastic differences in porosity evolution between the three materials.*

In conclusion, although all porous materials were subjected to the same macroscopic tensile axisymmetric loading corresponding to a constant macroscopic stress triaxiality $T_\Sigma$ =1 and constant $\mu_\Sigma = 1$ ($J_3^\Sigma > 0$) during the entire deformation process, the specificities of the plastic flow of the matrix, namely its SD ratio (which gives the sign of the material parameter k), dramatically affect the local state. All the results presented highlight the strong correlation between the sign of the macroscopic parameter k and the local plastic strain heterogeneity which leads to markedly different rates of void growth that ultimately strongly affect the ductility of the porous materials.

**4. Analysis of the porosity evolution and its effects on the ductility of porous materials under macroscopic axisymmetric tensile loading corresponding to $T_\Sigma = 1$ and $\mu_\Sigma = -1$**

Next, we examine the effect of the third-invariant $J_3$, on void growth. For this purpose, the macroscopic loading that is imposed corresponds to the same value of the stress triaxiality as in the previous case analyzed in Section 3, i.e. $T_\Sigma = 1$, but to opposite value of the Lode angle i.e. $\mu_\Sigma = -1$ ($J_3 < 0$ for the entire loading history). Specifically, the imposed macroscopic loading corresponds to the following ratios between the macroscopic principal true stresses : $\Sigma_1 = \Sigma_2$ and $\Sigma_3/\Sigma_1 = 0.25$.

Comparison between the macroscopic equivalent stress $\Sigma_e$ vs. macroscopic equivalent strain $E_e$ curves for the three materials is presented (Fig. 9). The evolution of the void volume fraction and the rate of void growth as a function of the macroscopic effective strain $E_e$, respectively are shown in Fig. 10-11.

Before analyzing in detail this new loading case, an important statement is required in order to drive the reader through the discussion of the simulation results. In the loading case previously analyzed, i.e. $T_\Sigma = 1$ and Lode parameter $\mu_\Sigma = +1$, the slowest rate of void growth was predicted for the porous material with matrix characterized by k = -0.3 (SD ratio $\sigma_T/\sigma_C < 1$). In contrast, for this loading which corresponds to $T_\Sigma = 1$ and $\mu_\Sigma = -1$, the slowest rate of void growth is now attained by the porous material characterized by k = +0.3 (SD ratio $\sigma_T/\sigma_C > 1$). Furthermore, this latter material shows enhanced ductility. This is consistent with the void volume fraction evolution (see Fig. 10), the rate of void growth in each material (Fig. 11) and the evolution of the relative inter-void ligament in the direction of the minimum applied stress shown in Fig. 12. Indeed, in the material with matrix characterized by k = -0.3, the void growth is much faster than in the other materials. Note that in all materials the relative inter-void ligament size in the z direction i.e. $l_3^r = l_3/l_3^0 = (C_3 - r_3)/(C_3^0 - r_0)$ decreases with the macroscopic effective strain $E_e$, the fastest decrease being in the material with k =-0.3. Moreover, in this material at some point in the deformation process, there is a rapid change in slope, which corresponds to the drop in the macroscopic effective stress (see Fig. 9). On the other hand, for the material with k = +0.3 there is a continuous decrease of $l_3^r$ (i.e. no change in slope) up to $E_e$ =1, when the calculations were stopped.

To gain understanding of the very large difference in the porosity evolution between the three materials, we also examine the distribution of the local equivalent plastic strain (Fig. 13) and mean stress (Fig. 14) in each material. The isocontours correspond to the same level of macroscopic effective strain $E_e$ = 0.20. It is very worth noting that in the material with the least plastic heterogeneity, which is the material with k = + 0.3, there is little damage (see also Fig 11). For this material, the relative mean stress $\sigma_m/Y_0$ distribution is also more homogeneous (see Fig. 14). In contrast, in the material with k = -0.3, the local plastic strain heterogeneity is much higher, the levels of mean stress and their gradients being also much higher than in the other cases.

While we have presented in detail the case of tensile loadings at $T_\Sigma$ =1, which shows the very strong link between the *sign of k and the rate of porosity evolution*, the same conclusions hold true for any moderate to high triaxiality $T_\Sigma > 0$. As an example, in Fig. 15 (a)-(b) is shown a comparison between the porosity evolution for the porous materials with matrix characterized by k =-0.3 and k = 0.3, respectively for macroscopic loadings corresponding to $T_\Sigma = 2$ and $T_\Sigma = 2/3$. Note that irrespective of the value of the stress triaxiality, for the material characterized by k =-0.3, the rate of void growth is much slower for tensile loadings with $\mu_\Sigma = 1$ ($J_3 > 0$ during the entire deformation process) than for tensile loadings with $\mu_\Sigma = -1$ ($J_3 < 0$ during the entire deformation process), while the reverse holds true for materials characterized by k =+0.3. At the same triaxiality, for loadings such that $\mu_\Sigma = 1$, the fastest rate of void growth occurs in the material characterized by k =+0.3, while for loadings such that $\mu_\Sigma = -1$, the fastest rate of void growth occurs in the material characterized by k = -0.3.

## 5. Conclusions

In this paper, the combined effects of the tension-compression asymmetry of the plastic flow of the incompressible matrix and third-invariant (or Lode parameter) on void growth were investigated. For this purpose, FE unit cell calculations were conducted. The plastic flow of the incompressible matrix was considered to obey the isotropic form of Cazacu et al. (2006)'s yield criterion. This yield criterion is pressure-insensitive, it involves all principal values of the Cauchy stress deviator, and a scalar material parameter, k. If k=0, the criterion reduces to von Mises. If k is different from zero, the criterion accounts for SD effects.

The imposed axisymmetric tensile loadings were such that the principal values of the macroscopic stresses, $\Sigma_1$, $\Sigma_2$, $\Sigma_3$ followed a prescribed proportional loading history corresponding to a specified value of the stress triaxiality. For the same triaxiality, calculations were carried out for loadings at either $J_3 > 0$ or $J_3 < 0$ (i.e. corresponding to the two possible values of the Lode parameter $\mu_\Sigma$).

It was clearly shown that irrespective of the imposed macroscopic loading, the tension-compression asymmetry in the plastic flow of the matrix, described by the parameter k, has a very strong influence on all aspects of the dilatational response of the porous solids. Furthermore, a very strong effect of the loading path, in particular of the Lode parameter on void evolution, and ultimately the material's ductility was observed. Specifically, for $\mu_\Sigma = +1$, the porous material with matrix characterized by k = -0.3 (yield in tension less than in compression: $\sigma_T/\sigma_C = 0.83$) has the highest ductility (~300% more than the material with k = + 0.3 for which $\sigma_T/\sigma_C = 1.21$). In contrast, for the porous von Mises material (k = 0; $\sigma_T/\sigma_C = 1$) and for the porous material characterized by k = + 0.3 damage is more gradual. Most importantly, although at the macroscopic level there is very little difference in the macroscopic stress-strain response between the porous materials, the differences in the local state fields are very pronounced. On the other hand, for the same triaxiality, but $\mu_\Sigma = -1$, the opposite holds true.

While only the case of macroscopic loadings at fixed stress triaxiality $T_\Sigma = 1$ was presented in detail, the following conclusions can be drawn concerning the dilatational response for porous materials subject to axisymmetric loadings corresponding to moderate to high values of positive stress triaxialities, namely that:

a) There is a very strong link between the sign of k (tension-compression asymmetry ratio of the matrix) and the rate of void growth;

(b) for the material characterized by k= +0.3, the rate of void growth is faster for $\mu_\Sigma = +1$ than for $\mu_\Sigma = -1$ while for the material characterized by k= -0.3, the opposite holds true.

It is also to be noted that the effect of the particularities of the plastic flow of the matrix, specifically the sign of k, on the yield surface of porous materials was established based on unit model cell calculations in Cazacu and Stewart (2009) and polycrystalline calculations in conjunction with the Fast Fourier Transform (FFT) method by Lebensohn and Cazacu (2012). For a porous material, whose matrix is softer in tension than in compression (k=-0.3), the yield points corresponding to axisymmetric states at $J_3 < 0$ are always above those for $J_3 > 0$. Meanwhile, for the porous material with k=-0.3, whose matrix is softer in compression than in tension, the opposite occurs i.e. the yield points that correspond to $J_3 > 0$ are above those for $J_3 < 0$. Since the plastic flow direction is along the normal to the yield locus, it means that the plastic flow and porosity evolution are strongly affected by the sign of k. And there is a clear correlation between the sign k and porosity evolution for $J_3>0$ and $J_3<0$, respectively.

While in the simulations presented in this paper, we studied only void growth and its influence on ductility in such materials, in view of industrial applications future studies devoted to the investigation of void collapse in such materials may provide valuable insights.

### References

1. Alves, J.L., Revil-Baudard, B., Cazacu, O. 2014. Importance of the coupling between the sign of the mean stress and the third-invariant on the rate of void growth and

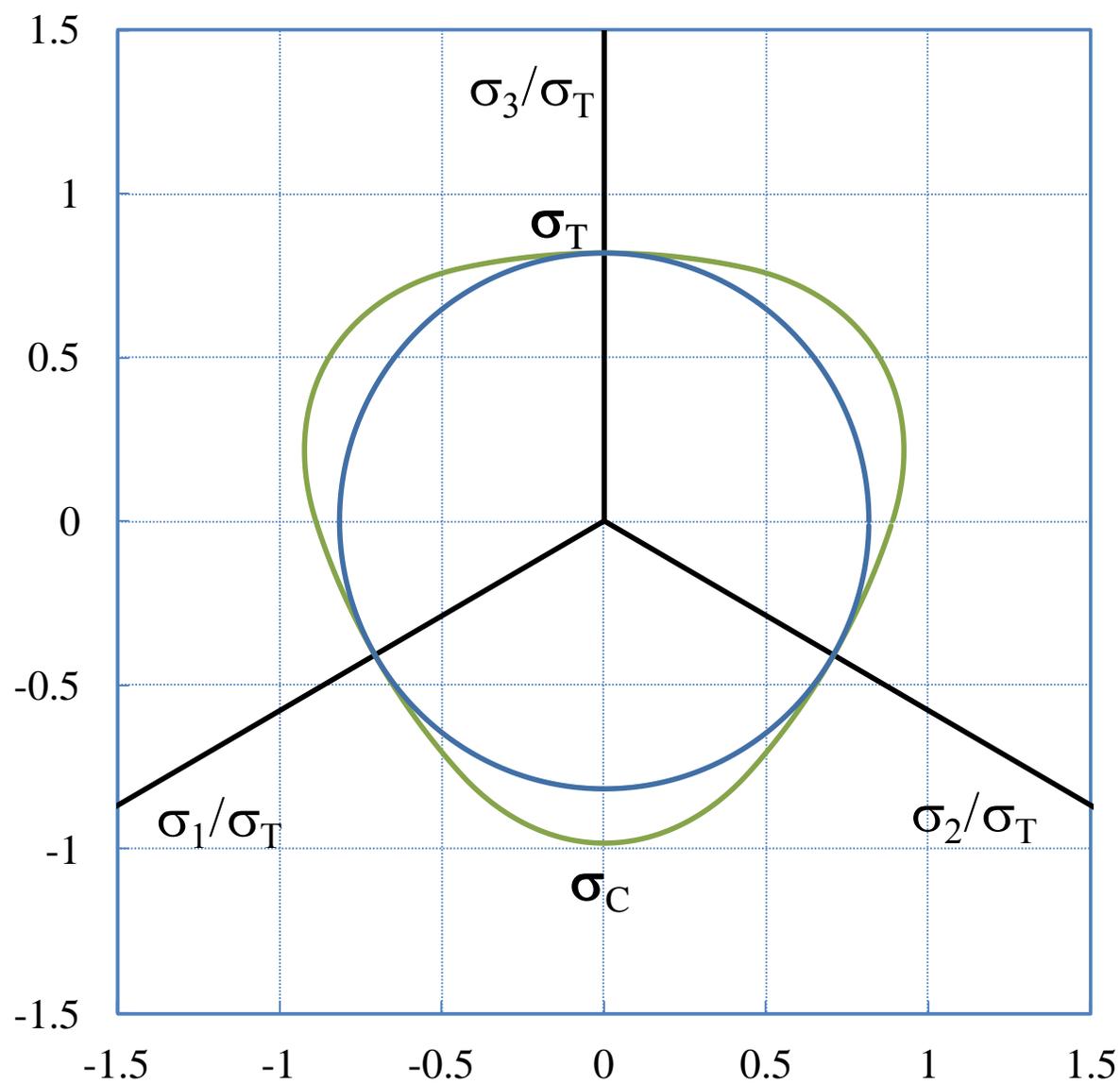

(a)

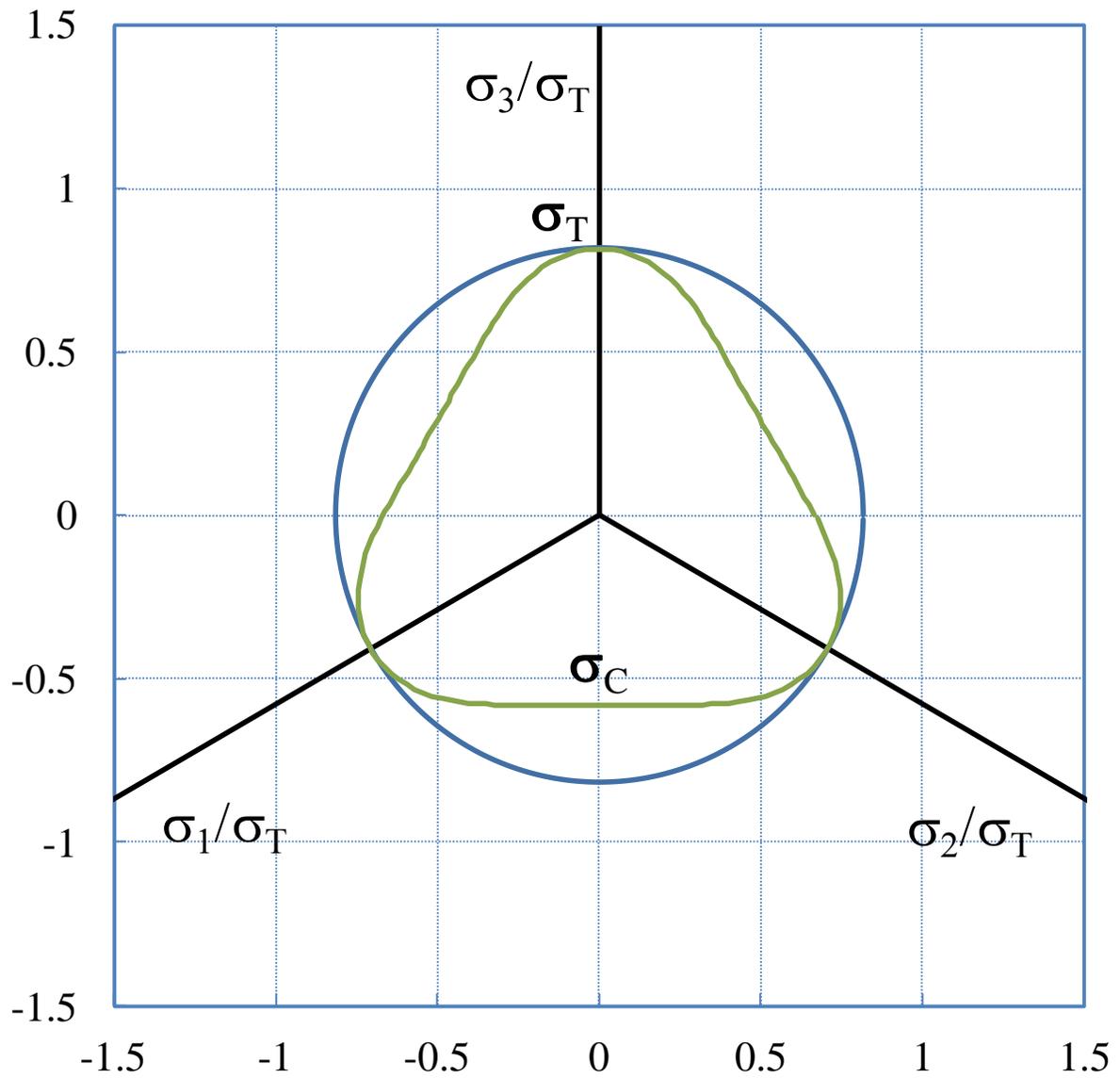

(b)

**Figure 1**: Representation in the octahedral plane of the yield locus according to the isotropic form of Cazacu et al. (2006) criterion for: (a) k = -0.3 ($\sigma_T/\sigma_C = 0.83$), and (b) k = + 0.3 ($\sigma_T/\sigma_C = 1.21$) in comparison with the von Mises criterion (circle, $\sigma_T/\sigma_C = 1$ and k =0).

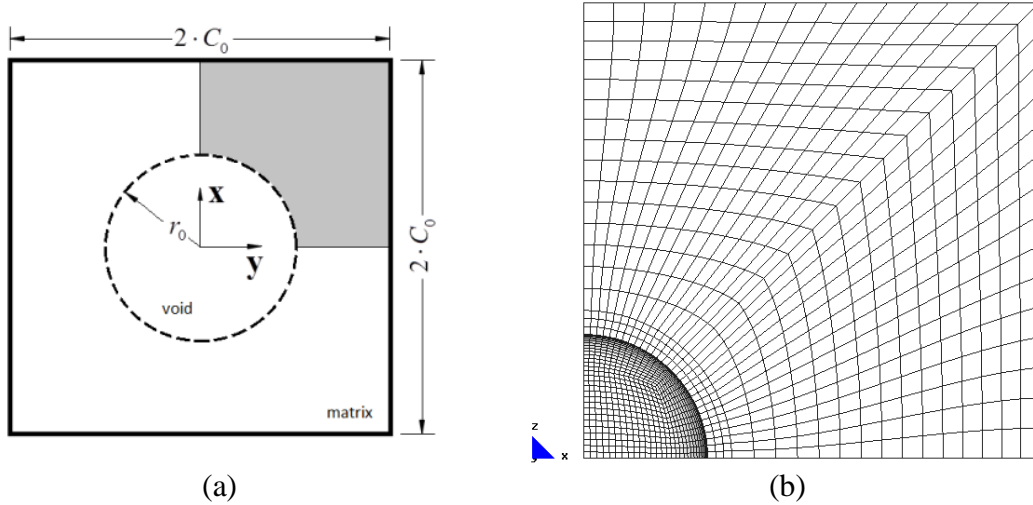

(a)                                                  (b)

**Figure 2:** (a) Schematic two-dimensional projection of the three-dimensional cubic cell model adopted in this study; 2 $C_0$ and $r_0$ denote the length of the undeformed cubic cell and the initial radius of the spherical void, respectively. (b) Finite-element mesh of one-eighth of the unit cell with a spherical void at its center.

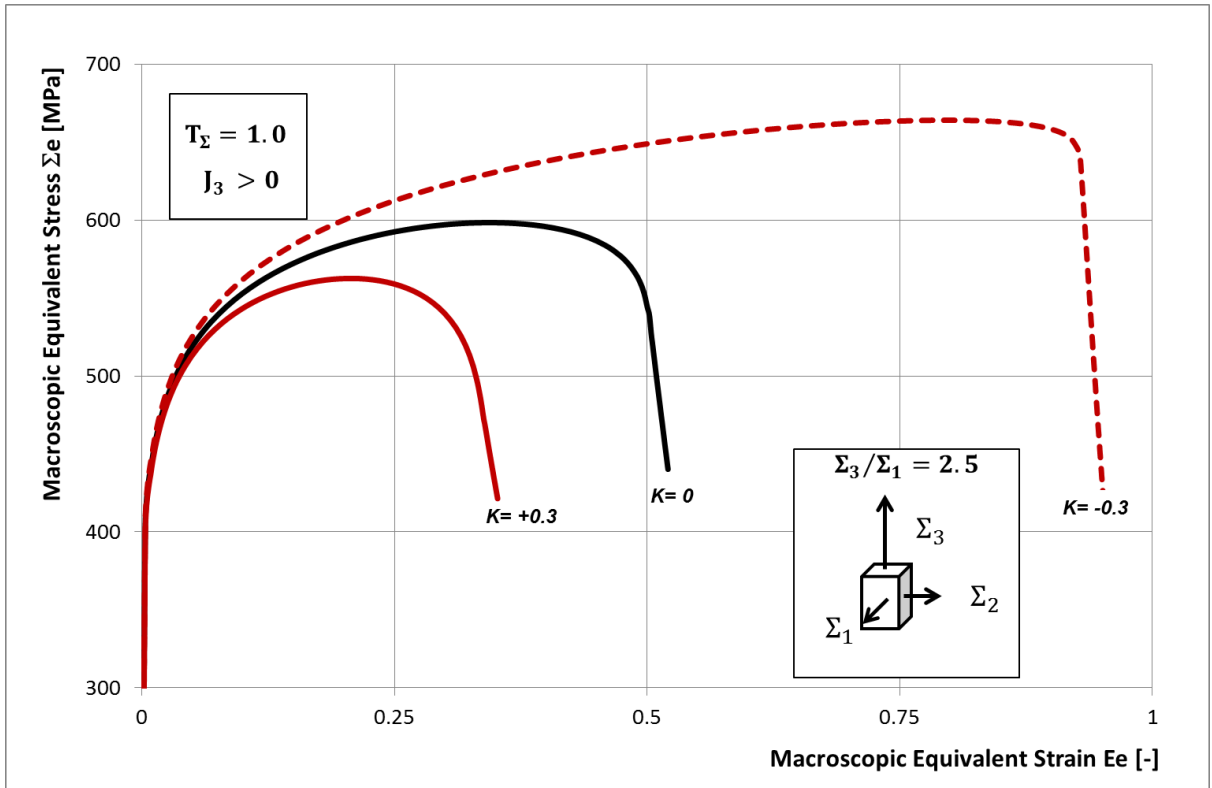

**Figure 3:** Comparison between the macroscopic stress-strain response for porous materials with matrix characterized by different tensile/compression asymmetry ratios: $\sigma_T/\sigma_C = 0.83$ ( k = - 0.3), von Mises material $\sigma_T/\sigma_C = 1$ (k = 0); $\sigma_T/\sigma_C = 1.21$ ( k = + 0.3) for axisymmetric loadings corresponding to a fixed triaxiality $T_\Sigma = 1$ and $J_3 > 0$ throught the deformation process ( Lode parameter $\mu_\Sigma = 1$).

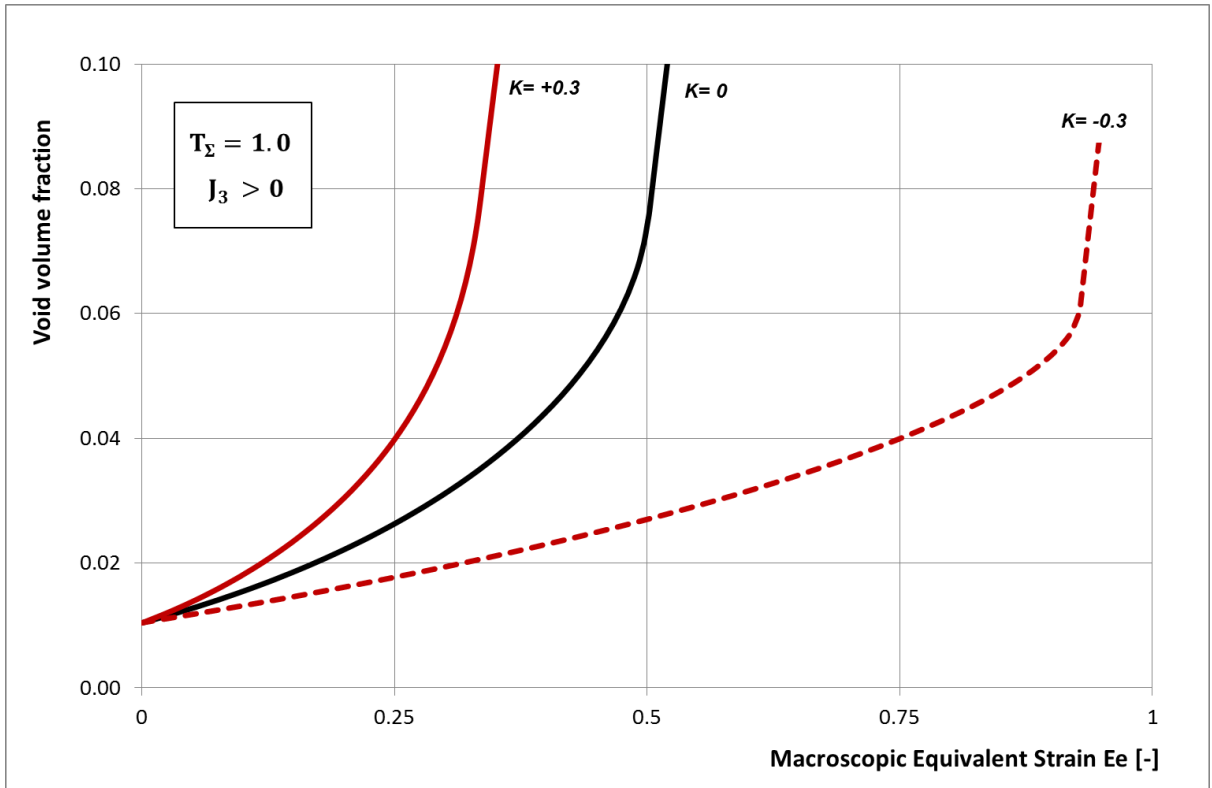

**Figure 4:** Evolution of the void volume fraction with the macroscopic equivalent strain $E_e$, for porous materials with matrix characterized by different tensile/compression asymmetry ratios: $\sigma_T/\sigma_C = 0.83$ ( k = - 0.3), von Mises material $\sigma_T/\sigma_C = 1$ (k = 0); $\sigma_T/\sigma_C = 1.21$ (k = + 0.3) for axisymmetric loadings corresponding to a fixed triaxiality $T_\Sigma = 1$ at $J_3 > 0$ ( Lode parameter $\mu_\Sigma = 1$).

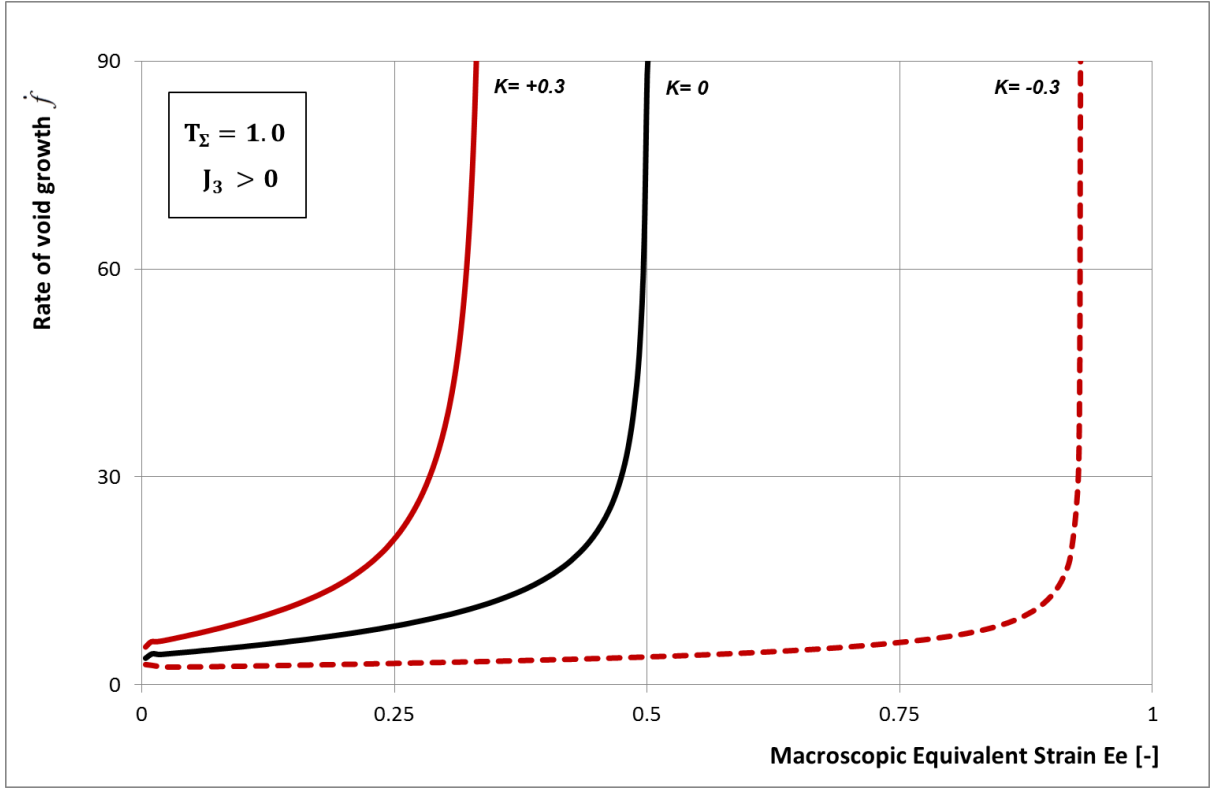

**Figure 5:** Evolution of the void growth rate ($\dot{f}$) with the macroscopic equivalent strain $E_e$, obtained by FE cell calculations for porous materials with matrix characterized by different tensile/compression asymmetry ratios: $\sigma_T/\sigma_C = 0.83$ ( k = - 0.3), von Mises material $\sigma_T/\sigma_C = 1$ (k = 0); $\sigma_T/\sigma_C = 1.21$ (k = + 0.3) for axisymmetric loadings corresponding to a fixed triaxiality $T_\Sigma = 1$ at $J_3 > 0$ ( Lode parameter $\mu_\Sigma = 1$).

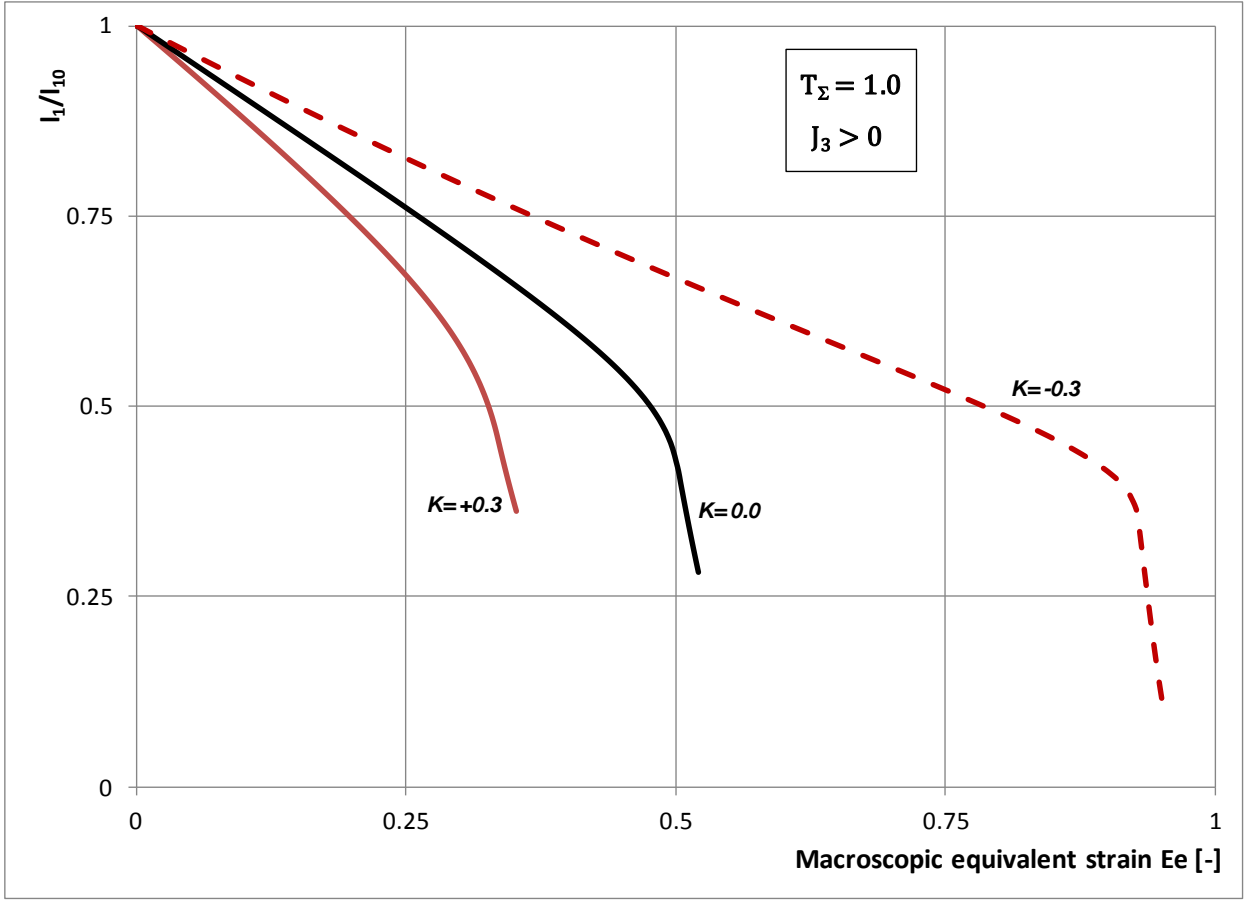

(a)

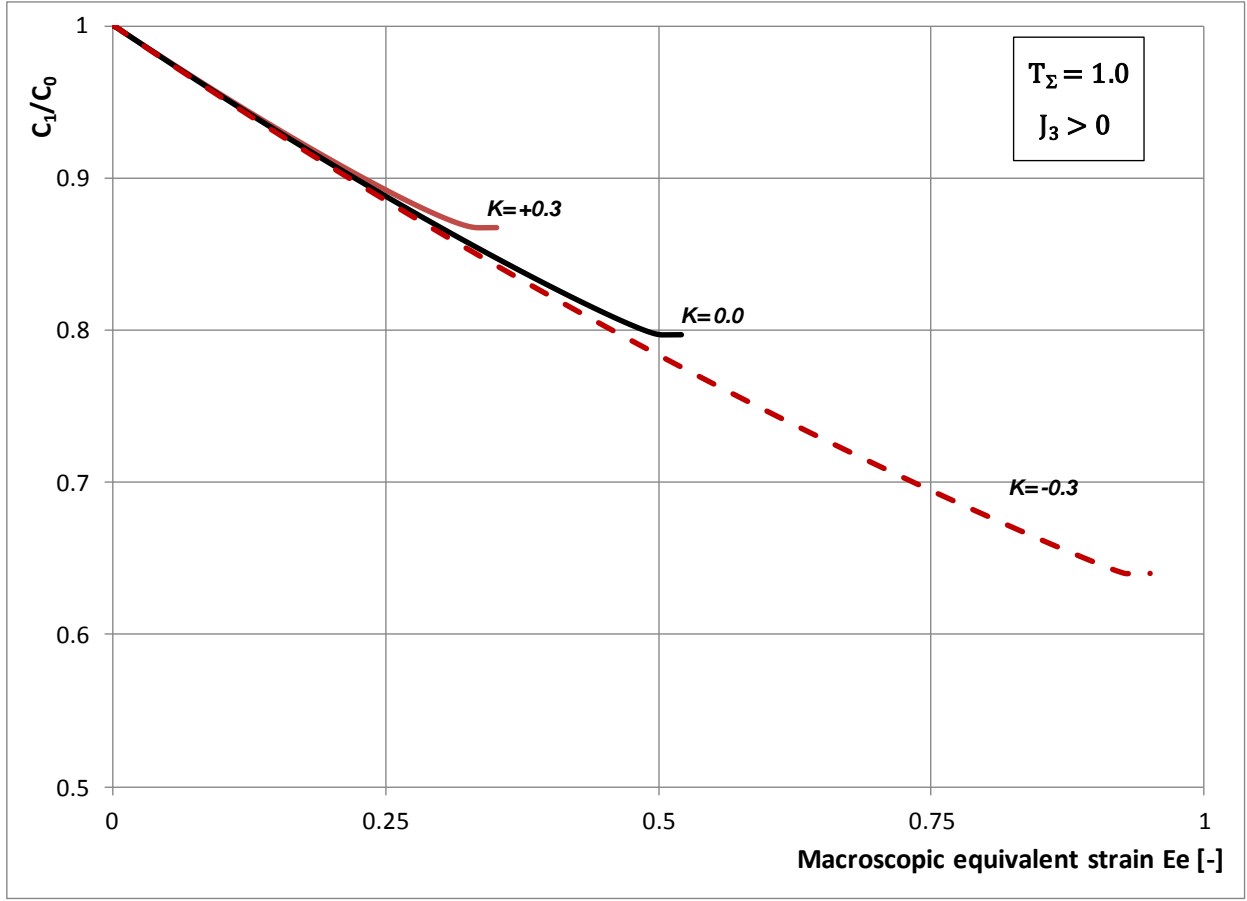

(b)

**Figure 6:** (a) Evolution of the inter-void ligament ratio $l_1^r = l_1/l_1^0$ in the direction of the minimum applied stress with the macroscopic equivalent strain $E_e$; and (b) evolution of the normalized unit cell dimension $C_1/C_0$ in the same direction for porous materials with matrix characterized by different tensile/compression asymmetry ratios: $\sigma_T/\sigma_C = 0.83$ ( k = - 0.3), von Mises material $\sigma_T/\sigma_C = 1$ (k = 0); $\sigma_T/\sigma_C = 1.21$  (k = + 0.3) for axisymmetric loadings corresponding to a fixed triaxiality $T_\Sigma = 1$ at $J_3 > 0$ ( Lode parameter $\mu_\Sigma = 1$).

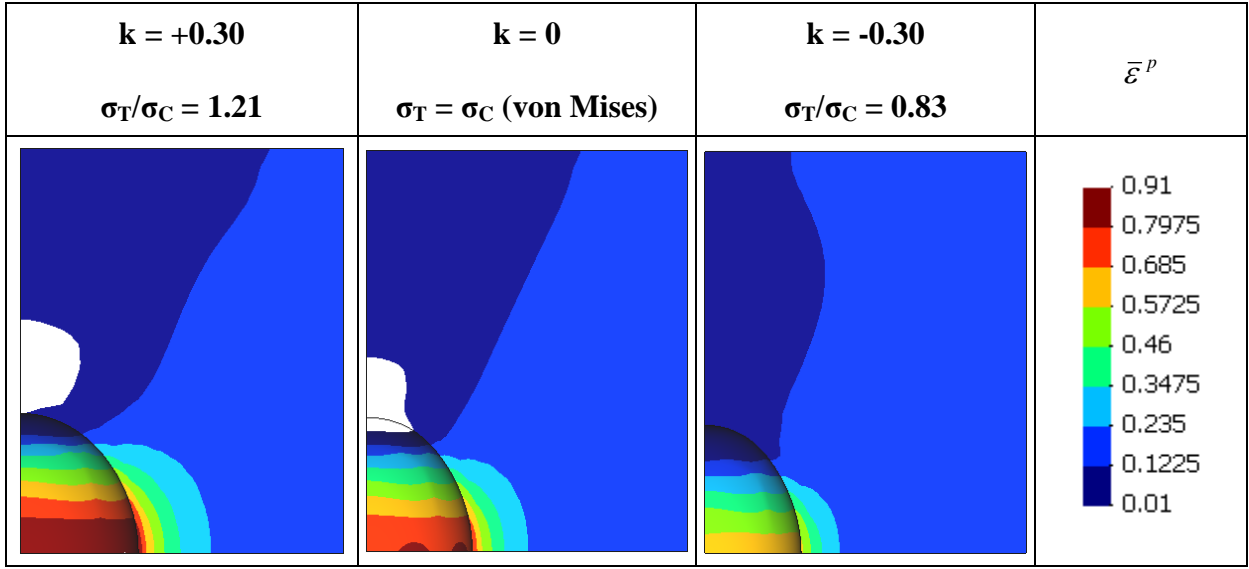

**Figure 7:** Isocontours of the local effective equivalent plastic strain $\bar{\varepsilon}^p$ corresponding to the same value of the macroscopic strain $E_e = 0.15$ for porous materials with matrix characterized by different tensile/compression asymmetry ratios: $\sigma_T/\sigma_C = 0.83$ ( k = - 0.3), von Mises material $\sigma_T/\sigma_C = 1$ (k = 0); $\sigma_T/\sigma_C = 1.21$ (k = + 0.3) for axisymmetric loadings corresponding to a fixed triaxiality $T_\Sigma = 1$ at $J_3 > 0$ ( Lode parameter $\mu_\Sigma = 1$). The white regions in the figure mark the elastic zones; *note that only in the material with k = -0.3, the entire domain is plastified*.

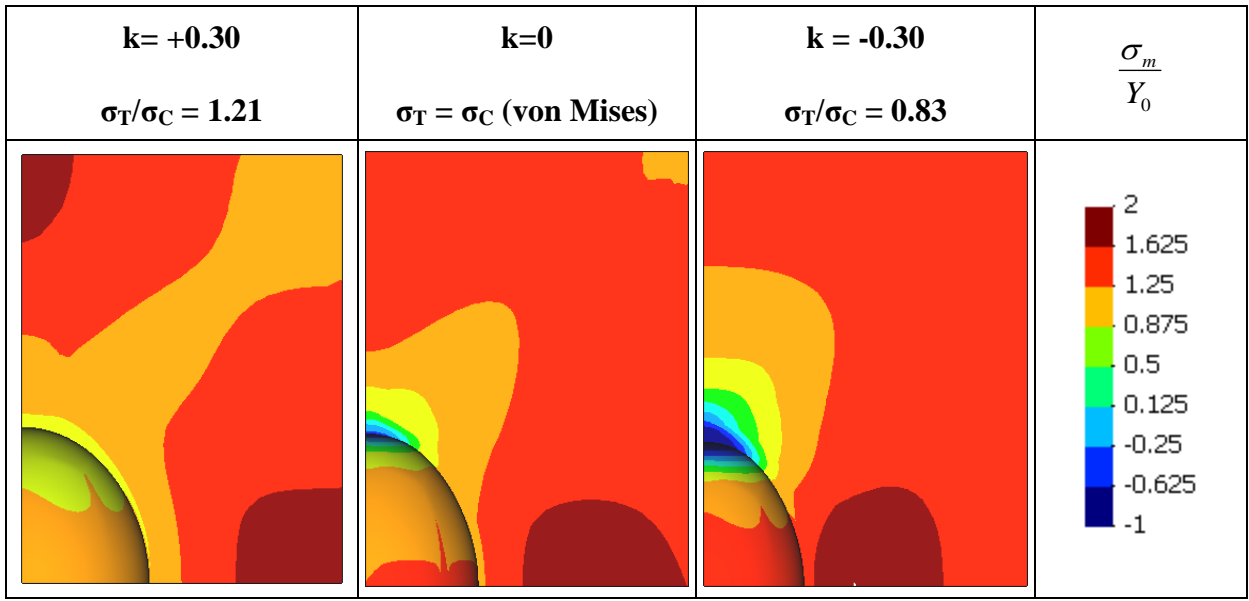

**Figure 8:** Isocontours of the local normalized mean stress $\sigma_m/Y_0$ for porous materials with matrix characterized by different tensile/compression asymmetry ratios: $\sigma_T/\sigma_C = 0.83$ ( k = - 0.3), von Mises material $\sigma_T/\sigma_C = 1$ (k = 0); $\sigma_T/\sigma_C = 1.21$ (k = + 0.3) for axisymmetric loadings corresponding to a fixed triaxiality $T_\Sigma = 1$ at $J_3 > 0$ ( Lode parameter $\mu_\Sigma = 1$).

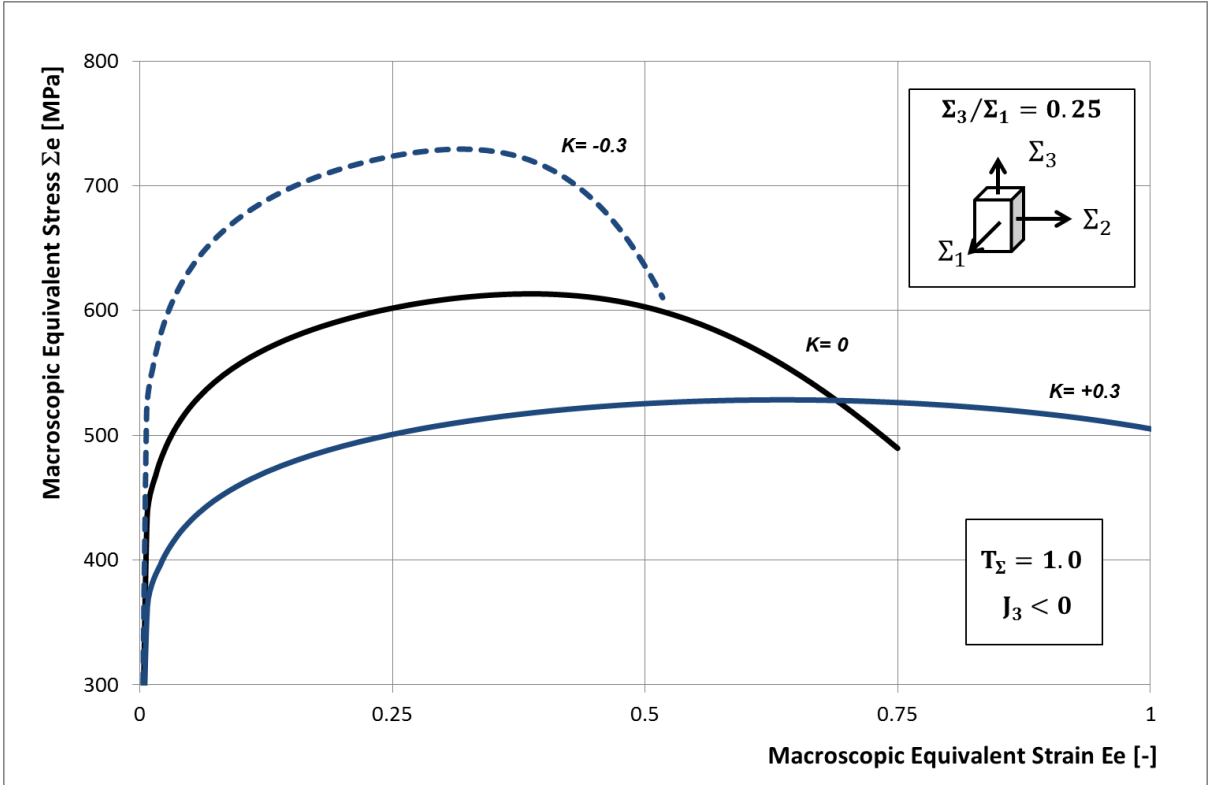

**Figure 9:** Comparison between the macroscopic stress-strain response obtained by FE cell calculations for porous materials with matrix characterized by different tensile/compression asymmetry ratios: $\sigma_T/\sigma_C = 0.83$ ( k = - 0.3), von Mises material $\sigma_T/\sigma_C = 1$ (k = 0); $\sigma_T/\sigma_C = 1.21$ (k = + 0.3) for axisymmetric loadings corresponding to a fixed triaxiality $T_\Sigma = 1$ at $J_3 < 0$ ( Lode parameter $\mu_\Sigma = -1$).

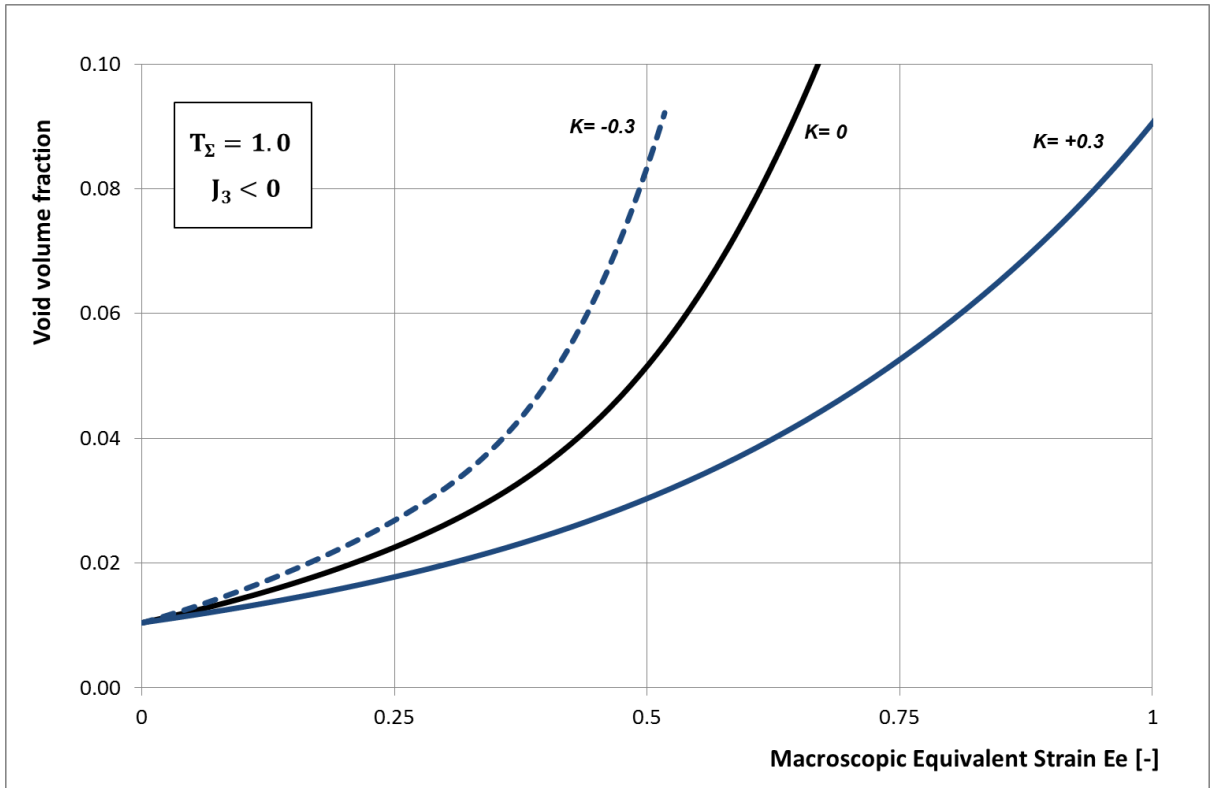

**Figure 10:** Evolution of the void volume fraction with the macroscopic equivalent strain $E_e$, obtained by FE cell calculations for porous materials with matrix characterized by different tensile/compression asymmetry ratios: $\sigma_T/\sigma_C = 0.83$ ( k = - 0.3), von Mises material $\sigma_T/\sigma_C = 1$ (k = 0); $\sigma_T/\sigma_C = 1.21$ (k = + 0.3) for axisymmetric loadings corresponding to a fixed triaxiality $T_\Sigma = 1$ at $J_3 < 0$ ( Lode parameter $\mu_\Sigma = -1$).

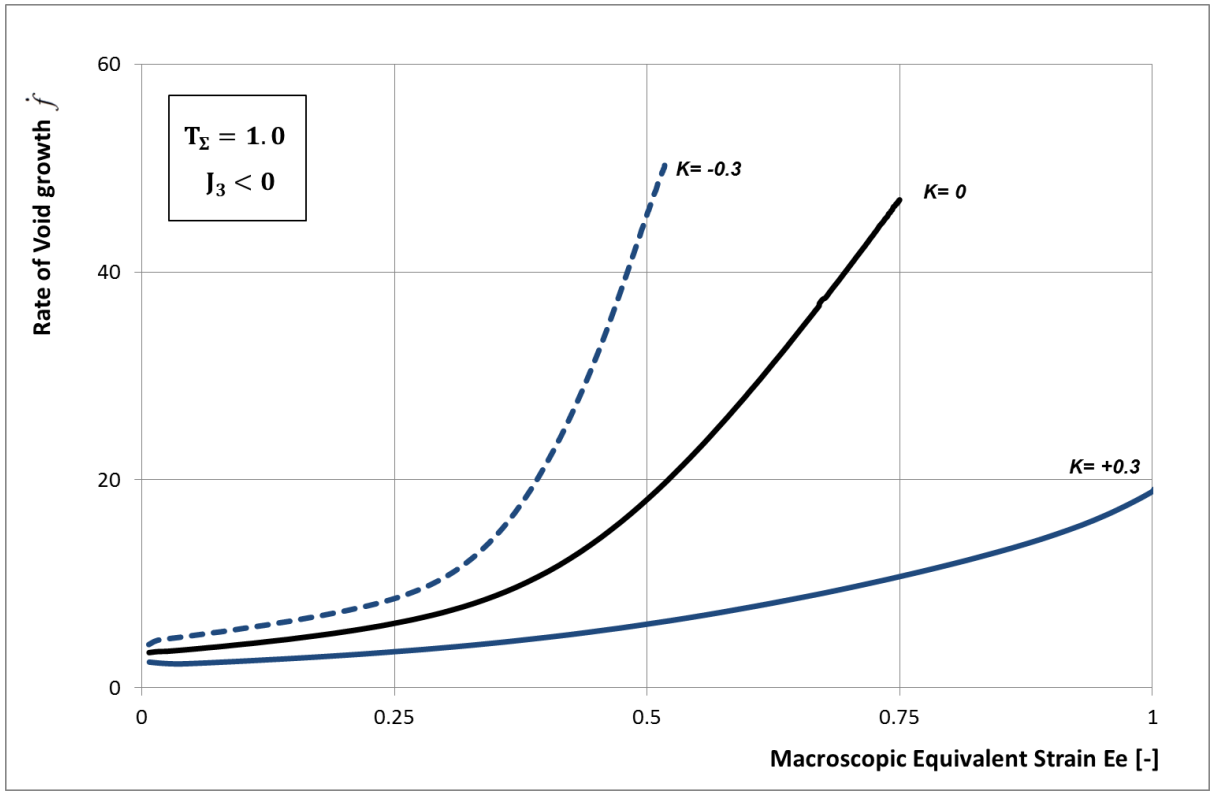

**Figure 11:** Evolution of the void growth rate ($\dot{f}$) with the macroscopic equivalent strain $E_e$, obtained by FE unit-cell calculations for porous materials with matrix characterized by different tensile/compression asymmetry ratios: $\sigma_T/\sigma_C = 0.83$ ( k = - 0.3), von Mises material $\sigma_T/\sigma_C = 1$ (k = 0); $\sigma_T/\sigma_C = 1.21$ (k = + 0.3) for axisymmetric loadings corresponding to a fixed triaxiality $T_\Sigma = 1$ at $J_3 < 0$ ( Lode parameter $\mu_\Sigma = -1$).

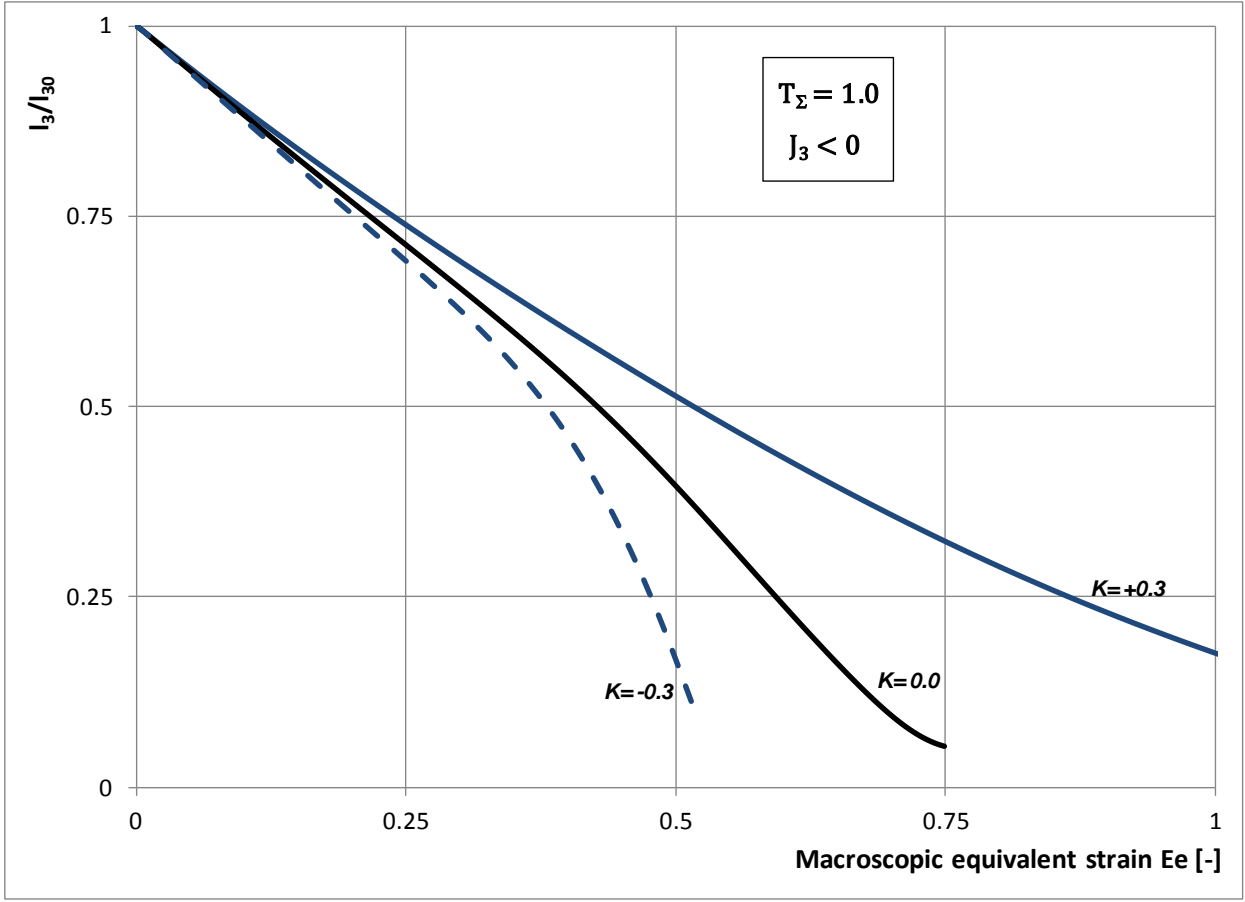

(a)

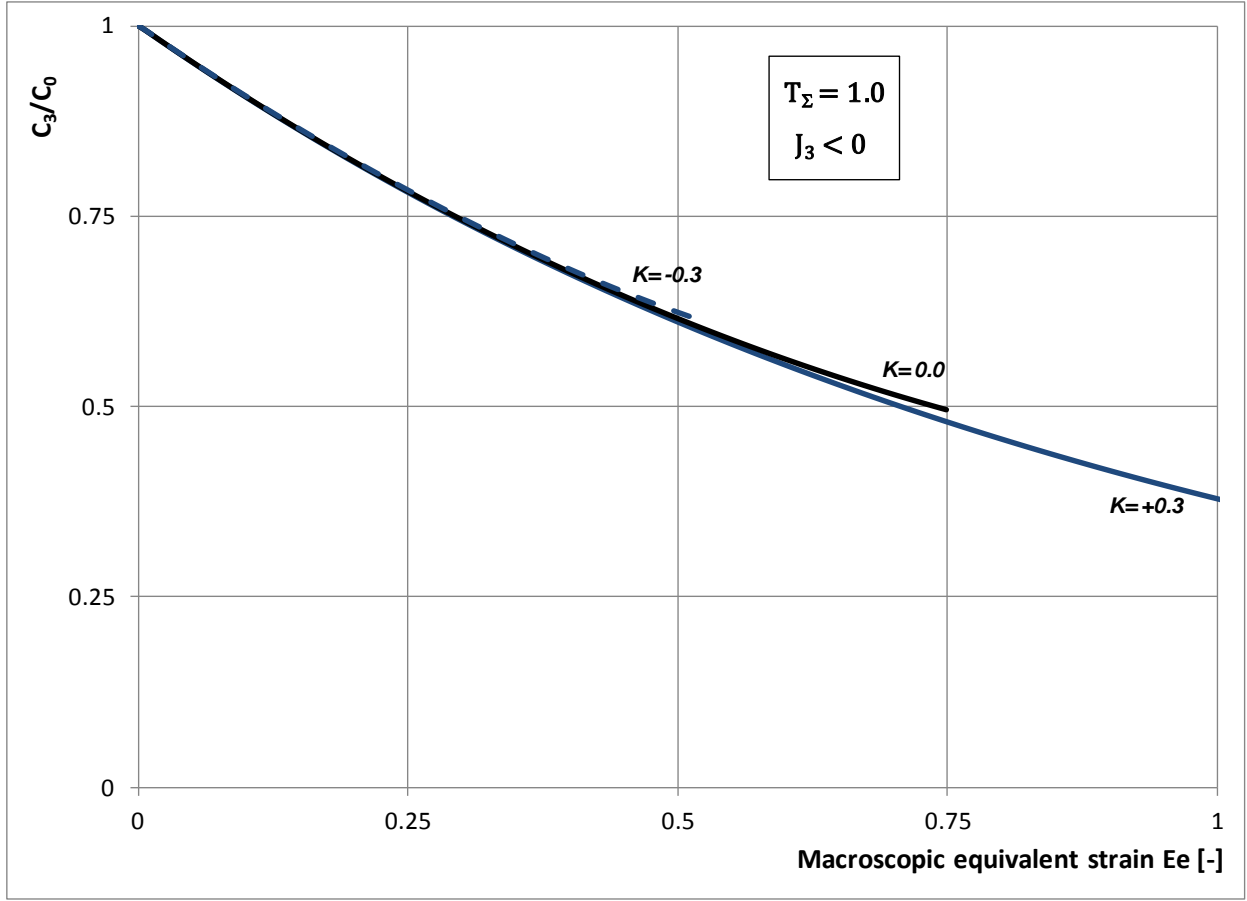

(b)

**Figure 12:** (a) Evolution of the relative inter-void ligament ratio in the direction of the minimum applied macroscopic load as a function of the macroscopic equivalent strain $E_e$; and (b) evolution of the normalized unit cell dimension $C_3/C_0$ in the same direction for porous materials with matrix characterized by different tensile/compression asymmetry ratios: $\sigma_T/\sigma_C = 0.83$ ( k = - 0.3), von Mises material $\sigma_T/\sigma_C = 1$ (k = 0); $\sigma_T/\sigma_C = 1.21$ (k = + 0.3) for axisymmetric loadings corresponding to a fixed triaxiality $T_\Sigma = 1$ at $J_3 < 0$ ( Lode parameter $\mu_\Sigma = -1$).

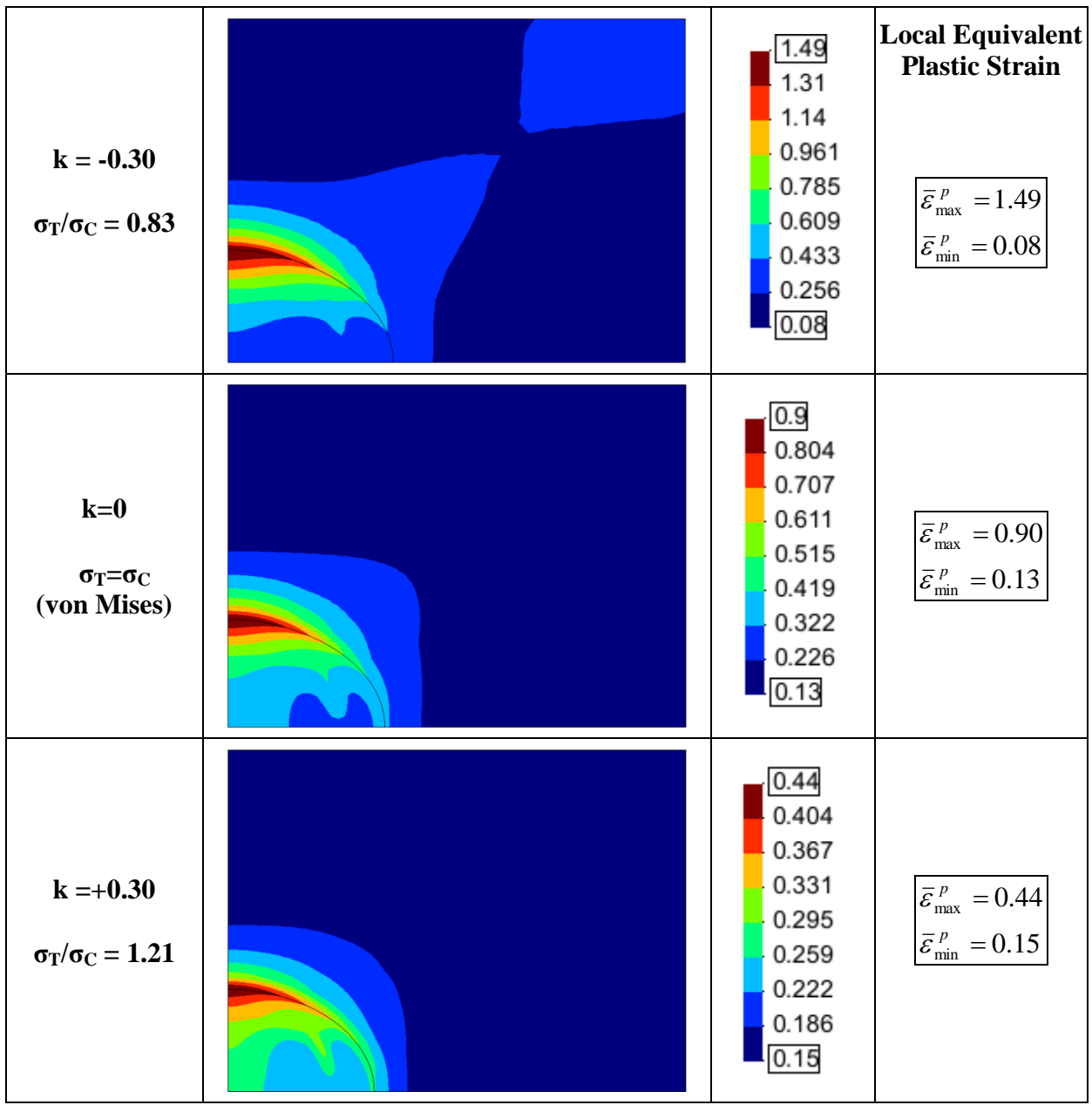

**Figure 13:** Isocontours of the local effective equivalent plastic strain $\bar{\varepsilon}^p$ corresponding to the same value of the macroscopic strain $E_e = 0.2$ for porous materials with matrix characterized by different tensile/compression asymmetry ratios: $\sigma_T/\sigma_C = 0.83$ ( k = -0.3), von Mises material $\sigma_T/\sigma_C = 1$ (k = 0); $\sigma_T/\sigma_C = 1.21$ ( k = + 0.3) for axisymmetric loadings corresponding to a fixed triaxiality $T_\Sigma = 1$ at $J_3 < 0$ ( Lode parameter $\mu_\Sigma = -1$). Vertical axial axis is along the Oz direction, horizontal axis along the Ox direction. Remark the huge differences between maximum and minimum values of the three cases analyzed.

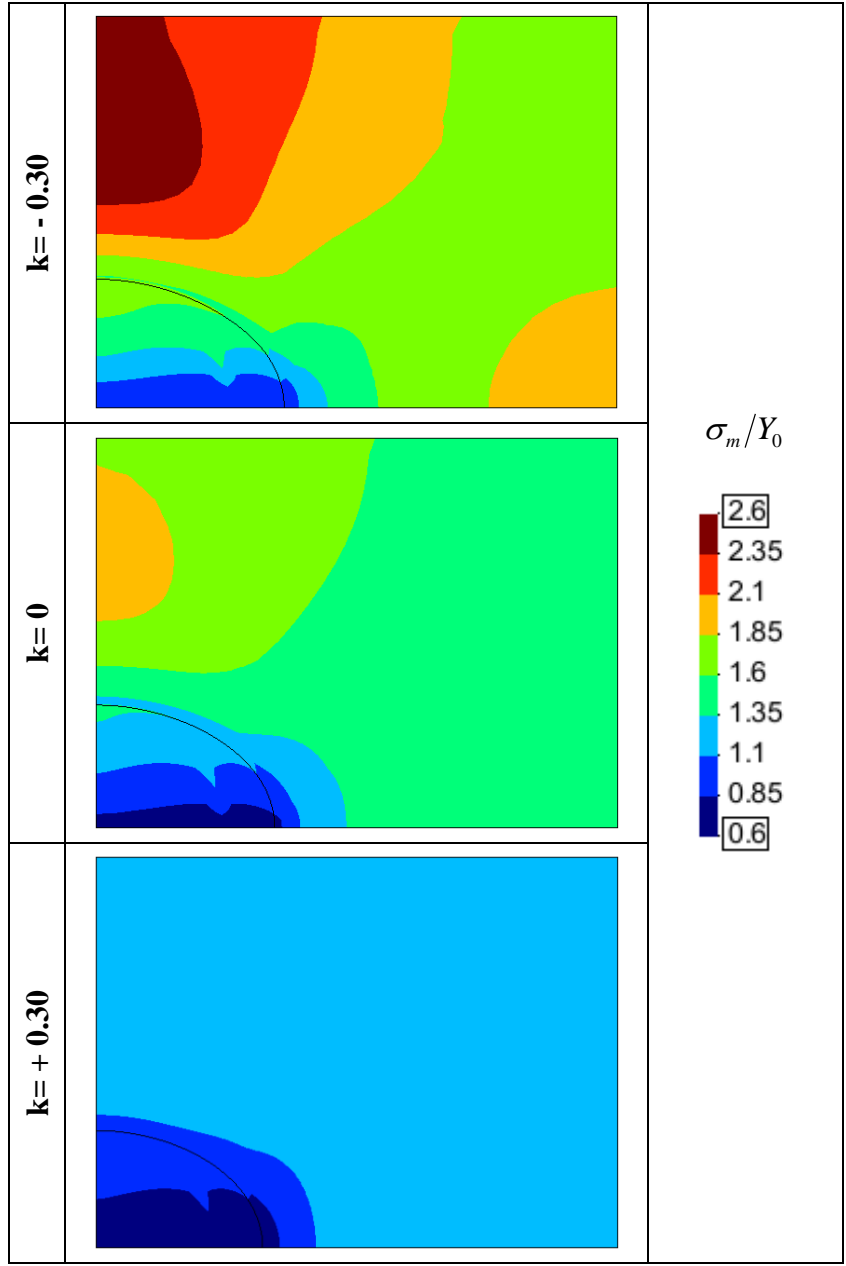

**Figure 14:** Isocontours of the normalized local means stress $\sigma_m/Y_0$, corresponding to the same value of the macroscopic strain $E_e = 0.2$ for porous materials with matrix characterized by different tensile/compression asymmetry ratios: $\sigma_T/\sigma_C = 0.83$ ( k = - 0.3), von Mises material $\sigma_T/\sigma_C = 1$ (k = 0); $\sigma_T/\sigma_C = 1.21$ (k = + 0.3) for axisymmetric loadings corresponding to a fixed triaxiality $T_\Sigma = 1$ at $J_3 < 0$ ( Lode parameter $\mu_\Sigma = -1$).

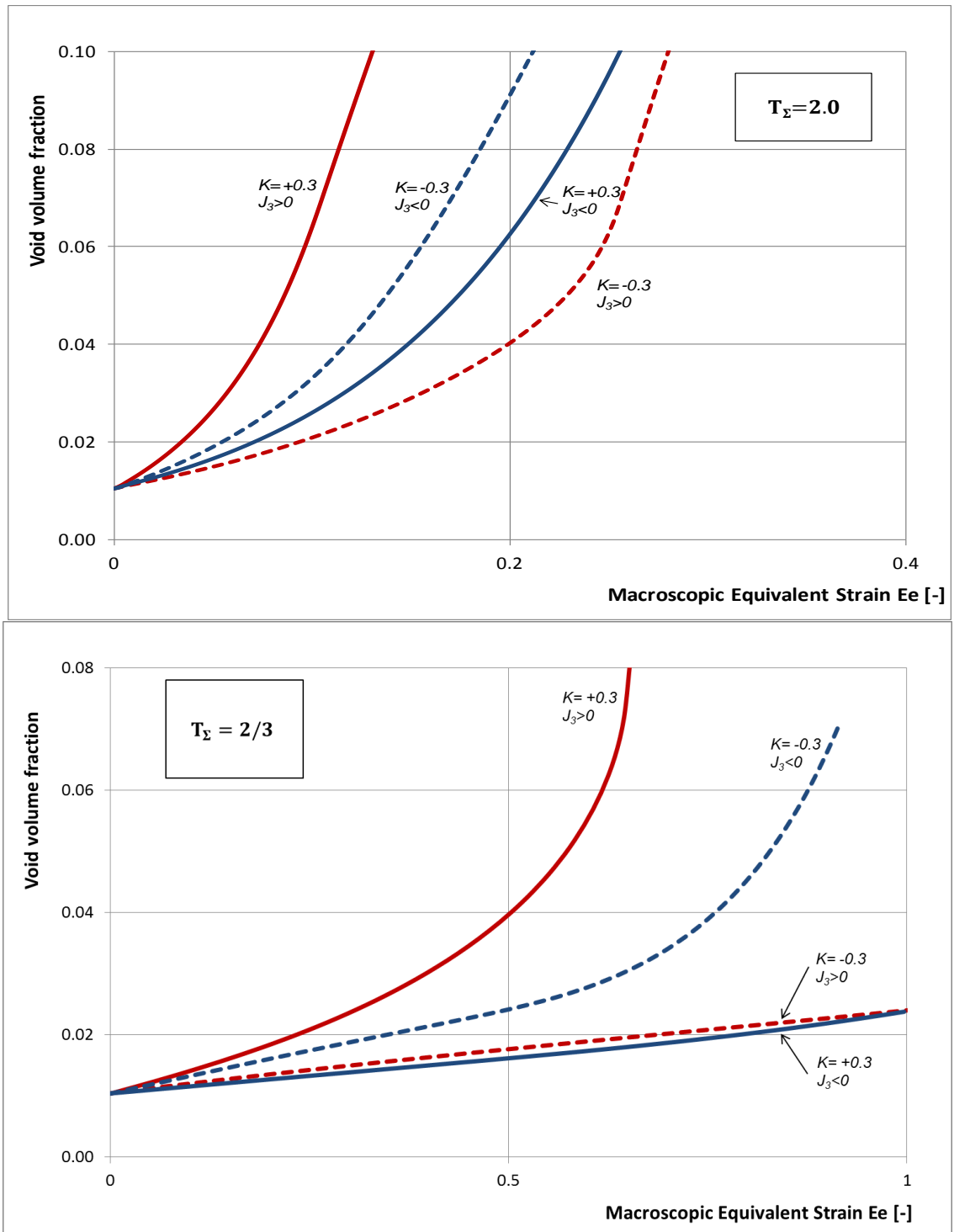

**Figure 15**: Porosity evolution for axisymmetric tensile loadings corresponding to fixed triaxiality (a) $T_\Sigma = 2$; (b) $T_\Sigma = 2/3$. Note that for loadings such that $J_3 > 0$ ( Lode parameter $\mu_\Sigma = 1$), void growth is fastest in the material with k = +0.3 while for loadings such that $J_3 < 0$ (Lode parameter $\mu_\Sigma = -1$), the fastest void growth occurs in the material with k = -0.3.